\documentclass[11pt]{amsart}

\usepackage{fullpage}
\usepackage{url}
\usepackage{amssymb}
\usepackage{cite}

\renewcommand{\a}{\alpha}
\renewcommand{\b}{\beta}
\newcommand{\g}{\gamma}
\renewcommand{\d}{\delta}
\newcommand{\D}{\Delta}
\newcommand{\e}{\varepsilon}
\newcommand{\f}{\varphi}

\renewcommand{\k}{\kappa}

\renewcommand{\t}{\theta}

\newcommand{\cM}{{\mathcal M}}

\newcommand{\cL}{{\mathcal L}}
\newcommand{\cE}{{\mathcal E}}
\newcommand{\cU}{{\mathcal U}}

\newcommand{\bR}{\mathbb R}

\newcommand{\bZ}{\mathbb Z}

\newcommand{\bE}{\mathbb E}

\newcommand{\be}{\begin{equation}}
\newcommand{\ee}{\end{equation}}

\newcommand{\bel}[1]{\begin{equation}\label{#1}}
\newcommand{\beaa}{\begin{eqnarray*}}
\newcommand{\bea}{\begin{eqnarray}}
\newcommand{\beal}[1]{\begin{eqnarray}\label{#1}}
\newcommand{\bean}{\begin{eqnarray}\nonumber}
\newcommand{\beadl}[1]{\begin{deqarr}\label{#1}}
\newcommand{\eeadl}[1]{\arrlabel{#1}\end{deqarr}}
\newcommand{\eeal}[1]{\label{#1}\end{eqnarray}}
\newcommand{\eead}[1]{\end{deqarr}}
\newcommand{\eea}{\end{eqnarray}}
\newcommand{\eeaa}{\end{eqnarray*}}

\renewcommand{\to}{\rightarrow}

\renewcommand{\phi}{\varphi}
\renewcommand{\epsilon}{\varepsilon}
\renewcommand{\hat}{\widehat}

\newcommand{\<}{\langle}
\renewcommand{\>}{\rangle}

\newcommand{\dm}{{\partial M}}

\theoremstyle{plain}
\newtheorem{theorem}{Theorem}[section]

\newtheorem{remark}[theorem]{Remark}

\newtheorem{corollary}[theorem]{Corollary}

\theoremstyle{definition}

\def\endproof{\qed \medskip}
\def\blacksquare{\hbox to .60em {\vrule width .60em height .60em}}

\numberwithin{equation}{section}

\date{\today}

\begin{document}

\title{Local Existence and Uniqueness for \\ exterior static vacuum Einstein metrics}

\author[ ]{Michael T. Anderson}
\address{Dept.~of Mathematics, Stony Brook University, Stony Brook, NY 11794-3651, USA} 
\email{anderson@math.sunysb.edu}
\urladdr{http://www.math.sunysb.edu/$\sim$anderson}

\thanks{Partially supported by NSF grant DMS 1205947 }

\begin{abstract}
We study solutions to the static vacuum Einstein equations on domains of the form 
$M \simeq \bR^{n+1}\setminus B$ with prescribed Bartnik data $(\g, H)$ on the 
inner boundary $\dm$. It is proved that for any smooth boundary data $(\g, H)$ close to standard 
round data on the unit sphere $(\g_{+1}, n)$, there exists a unique asymptotically flat 
solution of the static vacuum Einstein equations realizing the boundary data $(\g, H)$ which 
is close to the standard flat solution. 

\end{abstract}

\maketitle

\section{Introduction}\label{section:intro}
\setcounter{equation}{0}

  In this short note, we prove a local existence and uniqueness theorem for solutions of the static vacuum 
Einstein equations, motivated by a well-known conjecture of Bartnik in general relativity. To describe the 
result and its setting, recall that the static vacuum Einstein equations on an $(n+1)$-dimensional manifold 
$M$ are given by 
\be \label{stat}
uRic = D^{2}u, \ \ \D u = 0,
\ee
where $g$ is a Riemannian metric on $M$ and $u: M \to \bR^{+}$ is a positive (or non-negative) 
potential function. Here also $Ric$ is the Ricci curvature of $g$, $D^{2}$ is the Hessian and $\D = tr D^{2}$ 
is the Laplacian with respect to $g$. These equations are equivalent to the statement that the manifold 
$\cM = S^{1}\times M$, with metric 
$$g_{\cM} = u^{2}d\t^{2} + g,$$
is Ricci-flat, i.e.~$Ric_{g_{\cM}} = 0$, (the vacuum Einstein equations). In this work, we will only consider 
the situation where $M$ is diffeomorphic to the exterior of a ball $B \subset \bR^{n+1}$, so 
$M \simeq \bR^{n+1} \setminus B$. Moreover, we consider only solutions which are asymptotically 
flat (AF), in the sense that the metric $g$ and potential function $u$ satisfy
\be \label{af}
g - g_{Eucl} = 1 + O(r^{n-1}), \ \ u = 1 + O(r^{n-1}).
\ee
These conditions imply that the metric and potential in fact have an asymptotic expansion near infinity, 
cf.~\cite{BS}; this will not be relevant here however. 

  Let $\bE^{m,\a}$ be the space of all AF static vacuum solutions $(g, u)$ on $M$ which are $C^{m,\a}$ 
smooth up to the boundary $\dm$. We assume $m \geq 2$ and $\a \in (0,1)$. This space is given the 
$C^{m,\a}$ topology with weight function $r^{n-1-\e}$. Let ${\rm Diff}_{1}^{m+1,\a}$ be the group of 
AF diffeomorphisms of $M$ which equal the identity on $\dm$. This serves as the ``gauge" group of 
the problem and the quotient space 
\be \label{mod}
\cE^{m,\a} = \bE^{m,\a} / {\rm Diff}_{1}^{m+1,\a}
\ee
is the corresponding moduli space of AF static vacuum solutions. It is proved in \cite{A1}, 
cf.~also \cite{AK}, that $\cE^{m,\a}$ is a smooth infinite dimensional Banach manifold; in 
particular the group ${\rm Diff}_{1}^{m+1,\a}$  acts freely on $\bE^{m,\a}$. 
 
   With each $(g, u) \in \bE^{m,\a}$ one may associate its Bartnik boundary data $(\g, H)$, where $\g = 
g_{\dm}$ is the induced metric on $\dm$ and $H$ is the mean curvature with respect to the inward 
unit normal $N$ into $M$. Here the sign is chosen so that $H = n$ for the round unit sphere $S^{n}(1) \subset 
\bR^{n+1}$. This choice of boundary data was introduced by Bartnik \cite{Ba1} in connection with his definition 
of quasi-local mass in general relativity. Note that the boundary data are invariant under the gauge 
group ${\rm Diff}_{1}^{m+1,\a}$. 

   One may thus define a Bartnik boundary map 
\be \label{pi}
\Pi_{B}: \cE^{m,\a} \to Met^{m,\a}(\dm)\times C^{m-1,\a}(\dm),
\ee
$$\Pi_{B}(g, u) = (\g, H).$$
This is a map of Banach manifolds, and again it is proved in \cite{A1}, \cite{AK} that $\Pi_{B}$ is a smooth 
Fredholm map, of Fredholm index 0. This means that, at each $(g, u) \in \cE^{m,\a}$, the 
derivative or linearization
\be \label{dpi}
D\Pi_{B}: T\cE^{m,\a} \to T(Met^{m,\a})(\dm)\times C^{m-1,\a}(\dm),
\ee
$$D\Pi_{B}(h, u') = (h^{T}, H'_{h}),$$
is a Fredholm linear map, i.e.~has finite dimensional kernel and cokernel, with $dim Ker (D\Pi_{B}) = 
dim Coker(D\Pi_{B})$. Here $h$ is a symmetric bilinear form, $h^{T}$ is the restriction of $h$ to 
$T(\dm)$ and $H'_{h}$ is the variation of the mean curvature $H$ in the direction of the deformation 
$h$. The pair $(h, u')$ satisfy the linearization of the static vacuum equations \eqref{stat} and so give an  
infinitesimal Einstein deformation. 

  The pure gauge infinitesimal deformations are of the form $\k = (\cL_{Z}g, Z(u))$, where $Z$ is a 
$C^{m+1,\a}$ vector field on $M$ with $Z = 0$ on $\dm$. These are of course ``modded out" so 
don't appear in the finite dimensional kernel $K = Ker(D\Pi_{B})$. 

  A fundamental conjecture of Bartnik \cite{Ba1}, \cite{Ba2}, of importance in general relativity, is that the map $\Pi_{B}$ 
in \eqref{pi} is a global diffeomorphism, i.e.~one has global existence and uniqueness for this nonlinear boundary 
value problem, at least when the mean curvature $H > 0$. As discussed in \cite{AK}, this is not true in general 
and it remains a very interesting problem to understand the existence and uniqueness of solutions to the  
boundary value problem, i.e.~the injectivity and range of the map $\Pi_{B}$. This issue is also closely related to 
a deeper understanding of the Bartnik quasi-local mass, cf.~\cite{Ba1}, \cite{Ba2}. 

\medskip 

    Note that when the potential function satisfies $u = 1$ on $M$, then $M$ is necessarily flat, 
so that in the context above, $(M, g) \subset (\bR^{n+1}, g_{Eucl})$ isometrically. We will call the solution 
with $u = 1$, $g = g_{Eucl}$ and $M = M = \bR^{n+1} \setminus B^{3}(1)$, the {\em standard exterior 
solution}; thus $M$ is the exterior of the round unit ball in $\bR^{n+1}$ (up to isometry). The boundary 
data of the standard exterior solution are given by $(\g, H)$ with $\g = \g_{+1} = \g_{S^{n}(1)}$, 
and $H = n$.

\begin{theorem} The standard exterior of the round unit ball $M = \bR^{n+1} \setminus B^{3}(1)$, 
$g = g_{Eucl}$, $u = 1$ is a regular point of the boundary map $\Pi_{B}$, so that $D\Pi_{B}$ is an 
isomorphism. Thus 
\be \label{k}
K = Ker D\Pi_{B} = 0,
\ee
and $D\Pi_{B}$ is surjective at the standard exterior solution. 

  In particular, if $\k = (k, u') \in T\bE^{m,\a}$ is tangent to the standard exterior solution and 
$k^{T} = H'_{k} = 0$, then $k = \d^{*}Z$ with $Z = 0$ on $\dm = S^{n}(1)$. 
\end{theorem}

  A standard application of the inverse function theorem in Banach spaces gives the following:

\begin{corollary}
There is a neighborhood $\cU \subset Met^{m,\a}(S^{n})\times C^{m,\a}(S^{n})$ of the standard flat 
bounadry data $(\g_{+1}, n)$ such that for any $(\g, H) \in \cU$, there is a unique solution $(M, g, u)$ of the 
static vacuum Einstein equations \eqref{stat}, up to isometry in ${\rm Diff}_{1}^{m+1,\a}$, for which 
$$\Pi_{B}(g, u) = (\g, H).$$
\end{corollary}

  Corollary 1.2 generalizes a result of Miao \cite{M}, who proved this result when $dim M = 3$, for boundary 
data and static vacuum solutions which are invariant under reflection in the standard coordinate planes 
in $\bR^{3}$, i.e.~$\bZ_{2}\times \bZ_{2}\times \bZ_{2}$ invariant. Note that this condition does not 
allow one imply that the standard exterior solution is a regular point of $\Pi_{B}$. This is discussed 
further in Remark 2.1. Theorem 1.1 gives the first example of a regular point of $\Pi_{B}$, with the 
exception of solutions near to the exterior of the horizon of the Schwarzschild metric discussed in 
\cite{AK}. 

\medskip 

  I would like to thank Pengzi Miao for interesting discussions related to the topic of this paper. 
  
\section{Proofs and Remarks}
\setcounter{equation}{0}

  In this section we prove Theorem 1.1 and Corollary 1.2, and make several further remarks. The proofs 
are very simple, given the setting from the Introduction. To prove Theorem 1.1, it suffices to prove 
\eqref{k}, since $D\Pi_{B}$ is Fredholm, of Fredholm index 0. 

\medskip 

   We begin with the Gauss equation or scalar constraint equation for the boundary $\dm \subset M$. 
For a general Riemannian metric, this reads 
$$|A|^{2} - H^{2} + R_{\g} = R_{g} - 2Ric(N,N),$$
where $A$ is the 2nd fundamental form, $N$ is the unit normal pointing into 
$M$ and $R_{\g}$, $R_{g}$ are the scalar curvatures of $\g$ and $g$ respectively. For static solutions 
$R_{g} = 0$ and $-Ric(N,N) = -u^{-1}NN(u) = u^{-1}(\D_{\g}u + HN(u))$, where the second equality is a 
standard consequence of the relation $\D_{M}u = 0$ in adapted (Fermi) coordinates about $\dm$.  
(Here $\D_{\g}$ is the Laplacian on $(\dm, \g)$). Hence one has 
\be \label{1}
u(|A|^{2} - H^{2} + R_{\g}) = 2(\D_{\g} u + HN(u)),
\ee
Consider the linearization of \eqref{1} in the direction of an infinitesimal static Einstein deformation 
$\k = (k', u') \in K = Ker D\Pi$ about the standard round solution with $u = 1$. For the 
standard round boundary data on $S^{n}(1) \subset \bR^{n+1}$ one has $|A|^{2} = n$, $H^{2} = n^{2}$ 
and $R_{\g} = n(n-1)$, so that $|A|^{2} - H^{2} + R_{\g} = 0$.  We have $k^{T} = k|_{\dm} = 0$ and 
$A = \g$ so that $tr A'_{k} = H'_{k} = 0$. It follows that 
\be \label{2}
2(\D_{\g} u' + nN(u')) = 2\<A'_{k}, A\> - 2HH'_{k} = 2H'_{k} - 2nH'_{k} = 0,
\ee
on $\dm$. Hence  
$$\D_{\g} u' + nN(u') = 0,$$
on $\dm$. 

Consider in general the equation
\be \label{3}
\D_{\g} v + nN(v) = 0,
\ee
on $\dm$, where $v$ is harmonic on $(M, g_{Eucl})$ with $v \to 0$ at infinity. Note that the variation $u'$ 
satisfies these conditions, since 
\be \label{zero}
0 = (\D u)' = \D'u + \D u' = \D u',
\ee
again since $u = 1$ on $M$. Hence $u'$ is harmonic on $M$. It is clear from \eqref{af} that 
$u' \to 0$ at infinity in $M$. 

  We claim that the only solution to \eqref{3} is the trivial solution $v = 0$. Thus, suppose first that 
$\max v > 0$ at some point in $M$. Then $\max_{M} v > 0$ occurs at a point $p \in \dm$ and, 
by the Hopf maximum principle, cf.~\cite{GT} for instance, $-N(v(p)) > 0$ where $-N$ is the outward unit 
normal to $M$. Hence $\D v(p) \leq 0$ and $N(v(p)) < 0$, contradicting \eqref{3}. Hence $\max_{M}v  = 0$. 
The same argument working at a point achieving $\min v$ shows that $\min_{M}v = 0$ so that $v = 0$ 
which proves the claim. It follows in particular that $u' = 0$ on $M$. 

  By the linearized static vacuum equations, one has then $u'Ric + uRic' = (D^{2})'u + D^{2}u'$ which 
thus gives $Ric_{k}' = 0$, so that $k$ is an infinitesimal flat deformation of $(M \setminus B^{n+1}(1), g_{Eucl})$ 
with $k^{T} = 0$. It is well-known that the round sphere is infinitesimally isometrically rigid in $\bR^{n+1}$, 
i.e.~the only infinitesimal isometric deformations of $S^{n}(1)$ are trivial, pure gauge deformations. The 
proof briefly is as follows. The Gauss-Codazzi equations for a hypersurface $\dm$ in $\bR^{n+1}$ state 
$$d A = 0,$$
where $d$ is the exterior covariant derivative on vector-valued 1-forms. The linearization of this in the 
direction $k$ above gives 
$$d A'_{k} = 0,$$
so that $A'_{k}$ is a Codazzi tensor, with vanishing trace. It follows by [6, Thm.~16.9] for instance that 
$A'_{k} = 0$ when $\dm = S^{n}(1)$. Thus $k^{T} = A'_{k} = 0$. The linearized version of the fundamental 
theorem for hypersurfaces in $\bR^{n+1}$ then implies that $(k, u')$ is a pure gauge deformation so that 
$\k = (k, u')$ equals zero in $K = Ker(D\Pi)$. (Note here that Killing field deformations $X$ of $S^{n}(1)$ 
satisfy $k = \d^{*}X = 0$). 

{\endproof}

   To prove Corollary 1.2, the inverse function theorem in Banach spaces implies that $\Pi$ is a local 
diffeomorphism near the standard exterior solution. Corollary 1.2 is just a restatement of this fact. 

{\endproof}

  Another important class of static vacuum solutions are the Schwarzschild metrics of the
form
\be \label{Sch}
g_{m} = (1 - \frac{2m}{r^{n-1}})^{-1}dr^{2} + r^{2}g_{S^{n}(1)},
\ee
with $r > (2m)^{1/(n-1)}$ for $m \geq 0$ and $r > 0$ for $m \leq 0$, and 
\be \label{u}
u = \sqrt{1 - \frac{2m}{r^{n-1}}}.
\ee
Consider the exterior region $M = r^{-1}[1,\infty)$. One then has boundary data
$$\g = \g_{+1} = g_{S^{n}(1)}, \ \ {\rm and} \ \ H = n\sqrt{1 - 2m}.$$ 
The fact that regular points of the smooth map $\Pi_{B}$ are open implies that the exterior Schwarzschild 
domain $(M, g_{m}, u)$ above is a regular point of $\Pi_{B}$, for $m$ small. We conjecture that in fact all 
such round exterior Schwarzschild domains are regular points of $\Pi_{B}$. The proof of Theorem 1.1 
however does not go through in this case, mainly since the relation \eqref{zero} now becomes
$$\D u' = -\D' u = \<D^{2}u, k\> + \<\b(k), du\>,$$
where $\b = \d + \frac{1}{2}dtr$ is the Bianchi operator. It is not clear if the right side can be made to 
vanish, or have a particular sign, for $u$ of the form \eqref{u}. 

  It is also unknown if general flat exterior solutions are regular points of $\Pi_{B}$, i.e~$M = 
\bR^{n+1}\setminus B$ with $B$ a ball with smooth boundary $\partial B = \dm$, $g = g_{Eucl}$ 
$u = 1$. The main difficulty with the method above is handling the term $\<A'_{k}, A\>$. 
It would be very interesting to know this one way or the other. 

  We also conjecture that the standard flat exterior boundary data $(\g_{+1}, n)$, and similarly standard 
Schwarzschild exterior boundary data $(\g_{+1}, n\sqrt{1 - 2m})$ are uniquely realized by the flat and 
Schwarzschild metrics respectively. This is a global uniqueness conjecture, somewhat analogous to the 
well-known black hole uniqueness theorem for the Schwarzschild metric, cf.\cite{C}. It does not appear 
that the methods used above can be adapted to settle this issue.

\begin{remark}
{\rm It is of interest to compare the proof of Corollary 1.2 with the main result of Miao in \cite{M} which 
gives the same result in dimension 3 but with a $\bZ_{2}\times \bZ_{2}\times \bZ_{2}$ symmetry condition 
imposed. Loosely speaking, and in physics terminology, Miao works off-shell and with a fixed gauge and 
boundary conditions, while the approach in this work is on-shell (i.e.~on the space of solutions), without 
any gauge and studying the behavior of the gauge-invariant map to the boundary conditions. For this to 
work, it is important to know that the on-shell moduli space $\cE^{m,\a}$ is itself a smooth Banach manifold. 

  In a little more detail, Miao proves that the linearized Einstein operator in a Bianchi gauge $D\Phi$, 
together with the the boundary conditions $(\g, H)$ imposed, has a surjective linearization at the 
standard round exterior in $\bR^{3}$, when all data satisfy the symmetry condition 
$\bZ_{2}\times \bZ_{2}\times \bZ_{2}$ above. 

  Note that the operator $T \sim D\Phi$ in \cite{M} is not an isomorphism; the $L^{2}$ orthogonal 
complement of $Im T$ is described in [9, Prop.~2]. In particular, it has non-zero gauge components 
$\eta$, $\hat \eta$. The proof of Theorem 1.1 above implies that $Ker T = 0$ so that the Fredholm 
operator $T$ has negative Fredholm index. An examination of the form of $(Im T)^{\perp}$ given 
in [9, Prop.~2] also shows that $D\Pi_{B}$ in \eqref{dpi} is surjective in dimension 3, \cite{M2} so that 
the two methods of proof are consistent.

}
\end{remark}

\bibliographystyle{plain}

\end{document}